\documentclass[usenatbib, apjfonts]{emulateapj}

\usepackage{epsfig,subfigure}
\usepackage{amssymb}

\shorttitle{Effects of UVB and Stellar Radiation}
\shortauthors{Nagamine et al.}

\begin{document}

\title{Effects of Ultraviolet background and local stellar radiation on the \HI\ column density distribution}

\author{Kentaro Nagamine\altaffilmark{1,3}, Jun-Hwan Choi\altaffilmark{1,4},
Hidenobu Yajima\altaffilmark{2}}


\altaffiltext{1}{Department of Physics \& Astronomy, University of Nevada Las Vegas, 4505 Maryland Pkwy, Box 454002, Las Vegas, NV 89154-4002 USA; kn@physics.unlv.edu} 
\altaffiltext{2}{Department of Astronomy and Astrophysics,
Pennsylvania State University,
525 Davey Lab, University Park, PA 16802, USA}
\altaffiltext{3}{Visiting Researcher, Institute for the Physics and Mathematics of the Universe, University of Tokyo, 5-1-5 Kashiwanoha, Kashiwa, 277-8568, Japan}
\altaffiltext{4}{Current address: Department of Physics and Astronomy, University of Kentucky, Lexington, KY 40506-0055, USA}

\def\avg#1{\langle#1\rangle}
\newcommand{\Lam}{\Lambda}
\newcommand{\lam}{\lambda}
\newcommand{\Del}{\Delta}
\newcommand{\del}{\delta}
\newcommand{\mpc}{\rm Mpc}
\newcommand{\kpc}{\rm kpc}
\newcommand{\pc}{\rm pc}
\newcommand{\cm}{\rm cm}
\newcommand{\yr}{\rm yr}
\newcommand{\erg}{\rm erg}
\newcommand{\s}{\rm s}
\newcommand{\kms}{\,\rm km\, s^{-1}}
\newcommand{\Msun}{M_{\odot}}
\newcommand{\Lsun}{L_{\odot}}
\newcommand{\Zsun}{Z_{\odot}}
\newcommand{\hinv}{h^{-1}}
\newcommand{\himpc}{\hinv{\rm\,Mpc}}
\newcommand{\hikpc}{\hinv{\rm\,kpc}}
\newcommand{\himsun}{\,\hinv{\Msun}}

\newcommand{\Om}{\Omega_{\rm m}}
\newcommand{\Ol}{\Omega_{\Lam}}
\newcommand{\Ob}{\Omega_{\rm b}}
\newcommand{\OHI}{\Omega_{\rm HI}}
\newcommand{\HI}{H\,{\sc i}}
\newcommand{\HII}{H\,{\sc ii}}
\newcommand{\HeII}{He\,{\sc ii}}
\newcommand{\NHI}{N_{\rm HI}}
\newcommand{\fn}{f(\NHI)}
\newcommand{\Mstar}{M_{\star}}
\newcommand{\highz}{high-$z$}
\newcommand{\ngas}{n_{\rm gas}}
\newcommand{\nuv}{n_{\rm th}^{\rm UV}}
\newcommand{\nsf}{n_{\rm th}^{\rm SF}}
\newcommand{\chiHI}{\chi_{\rm HI}}


\begin{abstract}

We study the impact of ultraviolet background (UVB) radiation field 
and the local stellar radiation on the \HI\ column density distribution 
$\fn$ of damped Ly$\alpha$ systems (DLAs) and sub-DLAs at $z=3$ using 
cosmological smoothed particle hydrodynamics simulations.  
We find that, in the previous simulations with an optically thin 
approximation, the UVB was sinking into the \HI\ cloud too deeply, 
and therefore we underestimated the $\fn$ at $19<\log \NHI<21.2$ 
compared to the observations. 
If the UVB is shut off in the high-density regions with 
$\ngas > 6 \times 10^{-3}$\,cm$^{-3}$, then we reproduce the observed 
$\fn$ at $z=3$ very well. 
We also investigate the effect of local stellar radiation by post-processing 
our simulation with a radiative transfer code, and find that the local
stellar radiation does not change the $\fn$ very much. 
Our results show that the shape of $\fn$ is determined primarily by 
the UVB with a much weaker effect by the local stellar radiation 
and that the optically thin approximation often used in cosmological 
simulation is inadequate to properly treat the ionization structure 
of neutral gas in and out of DLAs. Our result also indicates that the DLA
gas is closely related to the transition region from optically-thick 
neutral gas to optically-thin ionized gas within dark matter halos.
\end{abstract}

\keywords{cosmology: theory --- galaxies: evolution --- galaxies: formation --- galaxies: high-redshift --- quasars: absorption lines --- methods: numerical}


\section{Introduction}
\label{section:intro}

The \HI\ column density distribution function $\fn$ is one of the 
most basic statistics of quasar absorption systems, similarly to 
the luminosity function of galaxies. 
The accuracy of observational data on $\fn$ has  
dramatically improved over the past several years, thanks to 
the large samples of damped Ly$\alpha$ systems (DLAs) 
and sub-DLAs discovered in large data sets of quasar spectra 
\citep[e.g.,][]{Peroux03, Pro04c, Peroux05, Pro05, Pro08c, Noterdaeme09}. 
Their results indicate that the observed $\fn$ of DLAs can be fitted well 
with either a double power-law or a Schechter-type function. 

It would be desirable to understand the physical origin of the shape of 
$\fn$ in a cosmological context of the standard $\Lam$ cold dark 
matter model. The first such attempt was made by \citet{Katz96b}, who used 
a cosmological smoothed particle hydrodynamics (SPH) simulation with a 
comoving box-size of 22.22$\himpc$, $2\times 64^3$ particles, 
and cosmological parameters 
$(\Omega_m, \Omega_{\Lam}, \Omega_b) = (1.0, 0.0, 0.05)$. 
They found that their simulation underpredicted $\fn$ compared to the 
observational data by a factor of few or more, using a uniform 
ultraviolet background (UVB) 
$J(\nu) = J_0 (\nu_0/\nu)$ at $z=3$, where $\nu_0$ is the 
Lyman-limit frequency and 
$J_0= 10^{-22}$\,erg\,s$^{-1}$\,cm$^{-2}$\,sr$^{-1}$\,Hz$^{-1}$.  
The UVB is usually treated with an optically thin limit 
regardless of gas density in cosmological hydrodynamic simulations 
owing to the computational limit. 
This simplified approximation may artificially increase 
the ionization fraction of gas, and gives rise to the discrepancy in $\fn$.

\citet{Nag04g, Nag07a} updated the work of \citet{Katz96b} using 
cosmological SPH simulations with a comoving box size of 
10$\himpc$, $2\times 324^3$ particles, and 
$(\Omega_m, \Omega_{\Lam}, \Omega_b)=(0.3, 0.7, 0.044)$.  
Interestingly, they also found a similar underprediction of $\fn$ 
compared to the observations, despite significantly higher 
resolution than that of \citet{Katz96b}. 
Their simulations included a uniform UVB of \citet{Haardt96} spectrum, 
modified by \citet{Dave99} to match the Ly$\alpha$ forest observations. 

We have attempted to resolve this discrepancy by modifying 
the models of star formation (SF) and supernova feedback; e.g., 
changing the SF threshold density, SF time-scale, feedback strengths, 
or adding metal-line cooling \citep{Choi09a, Choi09b}.  
However, none of these changes in the physical models resolved 
the discrepancy in $\fn$ fundamentally.

In this Letter, we show that the effect of UVB is the key in 
determining the shape of $\fn$ at $\log \NHI \lesssim 21.6$.
In addition, we consider the effect of local stellar radiation on $\fn$
by performing a radiative transfer (RT) calculation. 
Our paper is organized as follows. In Section~\ref{sec:sim}, we briefly 
describe the setup of our simulations, and present the results
in Section~\ref{sec:result}. We then discuss the comparison with other 
recent works, 
and conclude in Section~\ref{sec:discussion}.


\section{Simulations}
\label{sec:sim}

We use the updated version of the tree-particle-mesh SPH code 
{\small GADGET-3} \citep[originally described in][]{Springel05e}.
Our conventional code includes radiative cooling by H, He, and 
metals \citep{Choi09b}, heating by a uniform UVB of a modified 
\citet{Haardt96} spectrum \citep{Katz96a, Dave99}, SF, 
supernova feedback, a phenomenological model for galactic winds, 
and a sub-resolution model of multiphase interstellar medium 
\citep[ISM;][]{Springel03b}.
In this multiphase ISM model, high-density ISM is pictured to be 
a two-phase fluid consisting of cold clouds in pressure equilibrium 
with a hot ambient phase.
Cold clouds grow by radiative cooling out of the hot medium, and 
this material forms the reservoir of baryons available for SF.
We use the ``Pressure SF'' model described by \citet{Choi10a} (which is 
based on the work by \citet{Schaye08}), but we have checked that 
the details of the SF model do not change the main conclusions of 
this paper. 

For all the simulations used in this paper, we employ a box size of 
comoving 10\,$\himpc$ and a total particle number of $2\times 144^3$ 
for gas and dark matter. 
The initial gas particle mass is $m_{\rm gas}=4.1\times 10^6\,\himsun$, 
and the dark matter particle mass is $m_{\rm dm}=2.0\times 10^7\,\himsun$. 
The comoving gravitational softening length is $2.78\,\hikpc$, so the 
physical resolution of our simulation is $\sim$0.7\,$\hikpc$ at $z=3$. 
\citet{Nag04g} showed that increasing the particle number from 
$2\times 144^3$ to $2\times 324^3$ did not change the shape of $\fn$ 
very much, therefore our results would not be strongly affected 
by the resolution effect. (But see further discussion in 
Section~\ref{sec:discussion}.)
The comoving box size of 10\,$\himpc$ is somewhat small, however, 
the number of missed very massive haloes are relatively small, and 
the impact on $\fn$ is expected to be small.  In fact \citet{Nag04g} 
showed that $\fn$ did not change very much with increasing box size, 
except that the lower $\NHI$ end of $\fn$ decreased due to lower 
resolution.  In this work, we have not corrected our results for the 
box size effect. 
The adopted cosmological parameters of all simulations are consistent 
with the latest WMAP result \citep{Komatsu09, Komatsu10}: 
$(\Om,\Ol,\Ob,\sigma_8, h, n_s)= (0.26, 0.74, 0.044, 0.80, 0.72, 0.96)$, 
where $h=H_0 / (100\kms\,\mpc^{-1})$.

With this setup, we run four simulations with different models of UVB: 
``Fiducial'', ``No-UV'', ``Half-UV'', and ``OTUV'' (Optically Thick UV) runs.
In the Fiducial run, the gas is heated and ionized by the uniform UVB 
under the optically thin approximation. 
In the No-UV run, the UVB strength is set to zero. 
In the Half-UV run, the normalization of UVB is reduced by half.
In the OTUV run, we assume that the uniform UVB cannot penetrate 
into the high-density gas with $\ngas > \nuv$, but otherwise it is 
the same as the Fiducial run at $\ngas \le \nuv$.

We adopt the threshold density $\nuv = 0.01 \, \nsf = 6\times 10^{-3}$\,cm$^{-3}$, where $\nsf$ is the SF threshold density above which 
the stars are allowed to form.  In our simulations, the gas with 
$\ngas > \nsf$ is mostly neutral owing to the multiphase ISM model. 
We originally arrived at the above value of $\nuv$ by successively 
lowering its value from $\nsf$ and checking the agreement with 
the observed $\fn$, but will provide further justifications below. 

There is evidence that the above value of $\nuv$ is physically appropriate. 
\citet{Tajiri98} found that the hydrogen cloud becomes fully 
self-shielded above a critical density of $1.4\times 10^{-2}$\,cm$^{-3}$ 
through RT calculations for a spherical top-hat sphere, and that 
the critical density has a mild dependence on the cloud mass and 
the UVB intensity.  
\citet{Kollmeier10} performed a three-dimensional UVB RT calculation with an 
isothermal sphere, and showed that the above value of $\nuv$ 
approximately corresponds to the transition density from \HII\ to \HI.
\citet{Faucher10} postprocessed cosmological SPH simulations with 
a ray tracing code and found that turning off UVB at 
$\ngas > 0.01$\,cm$^{-3}$ produces a favorable result. 
Furthermore, we also confirmed that the above $\nuv$ is appropriate 
by postprocessing our simulations with a RT code, which we will report 
in detail in a separate paper (H. Yajima et al., 2011, in preparation). 
For these reasons, we consider that the correct value of $\nuv$ is 
in the range of $10^{-2}$ to $10^{-3}$\,cm$^{-3}$, depending on the 
cloud mass and UVB intensity. 

\begin{figure}
\begin{center}
\includegraphics[angle=0, scale=0.44]{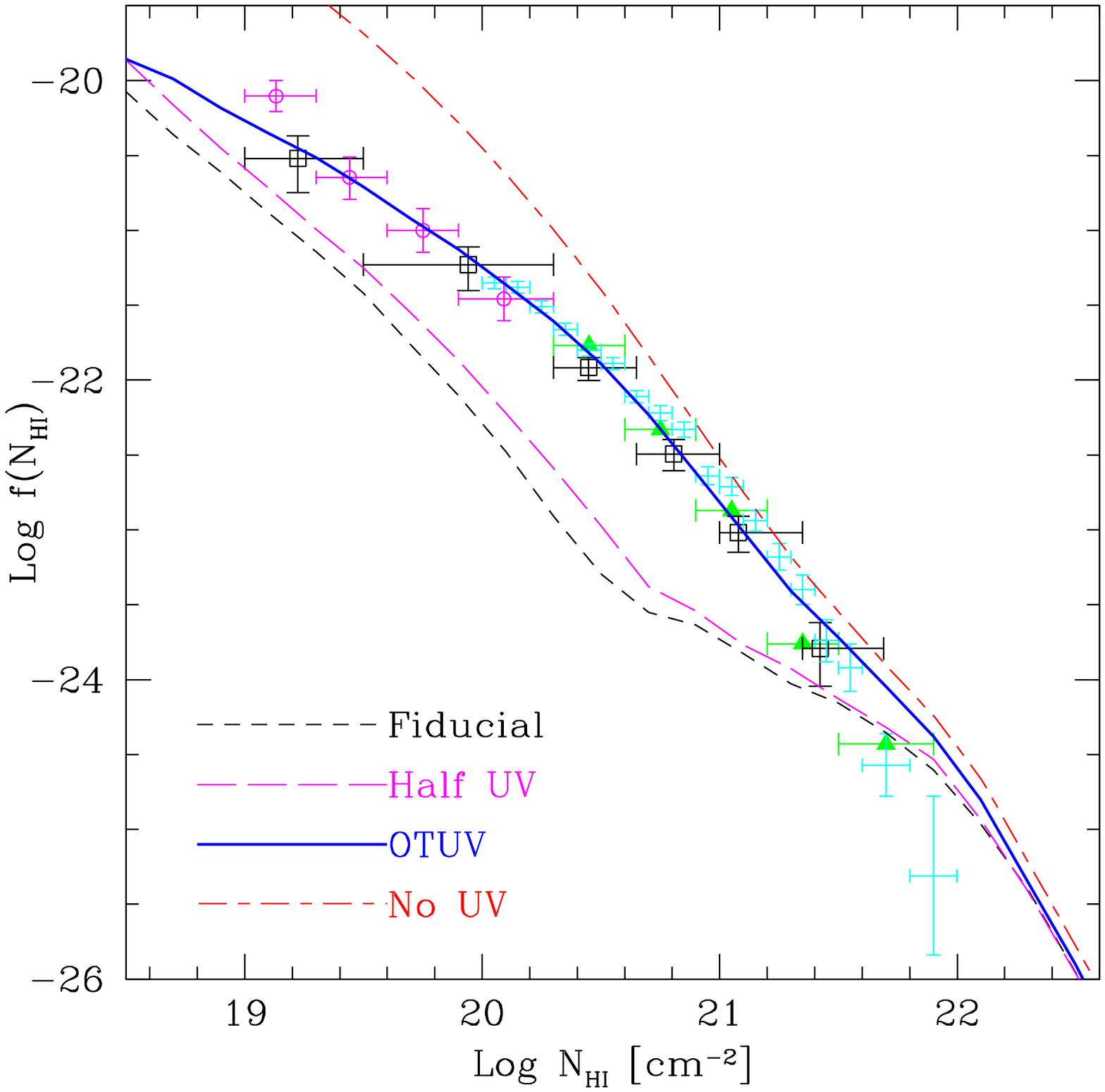}
\caption{\HI\ column density distribution functions 
at $z=3$ for the four runs with different treatment of UVB.
The observational data points are from 
\citet[][black open squares]{Peroux05}, 
\citet[][magenta open circles]{Omeara07},
\citet[][green triangles]{Pro09}, and 
\citet[][cyan bars]{Noterdaeme09}.
}
\label{fig:fn}
\end{center}
\end{figure}

\begin{figure*}
\centerline{
 \subfigure[Fiducial]{\includegraphics[width=0.225\textwidth,angle=0] {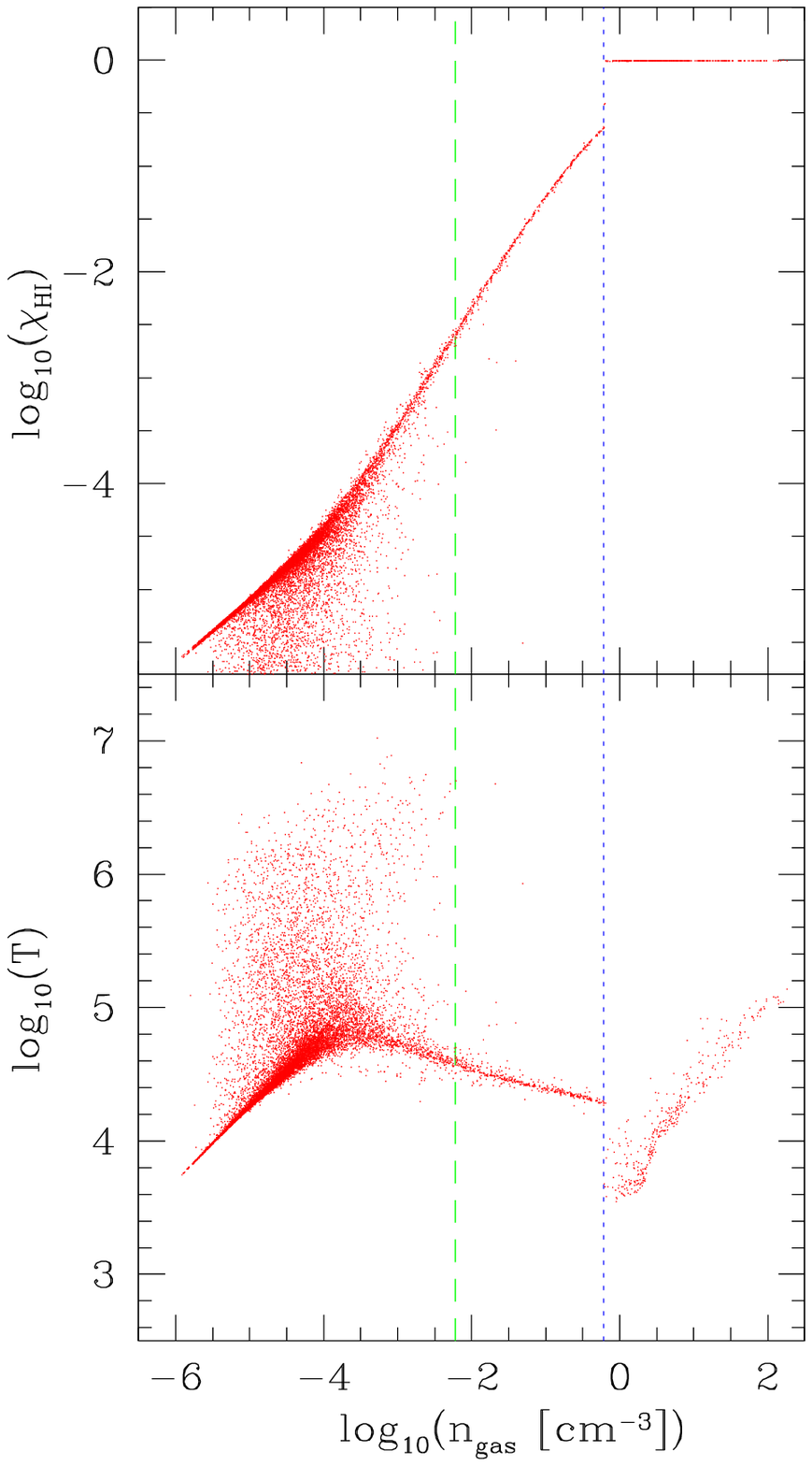}}
  \subfigure[No-UV]{\includegraphics[width=0.225\textwidth,angle=0] {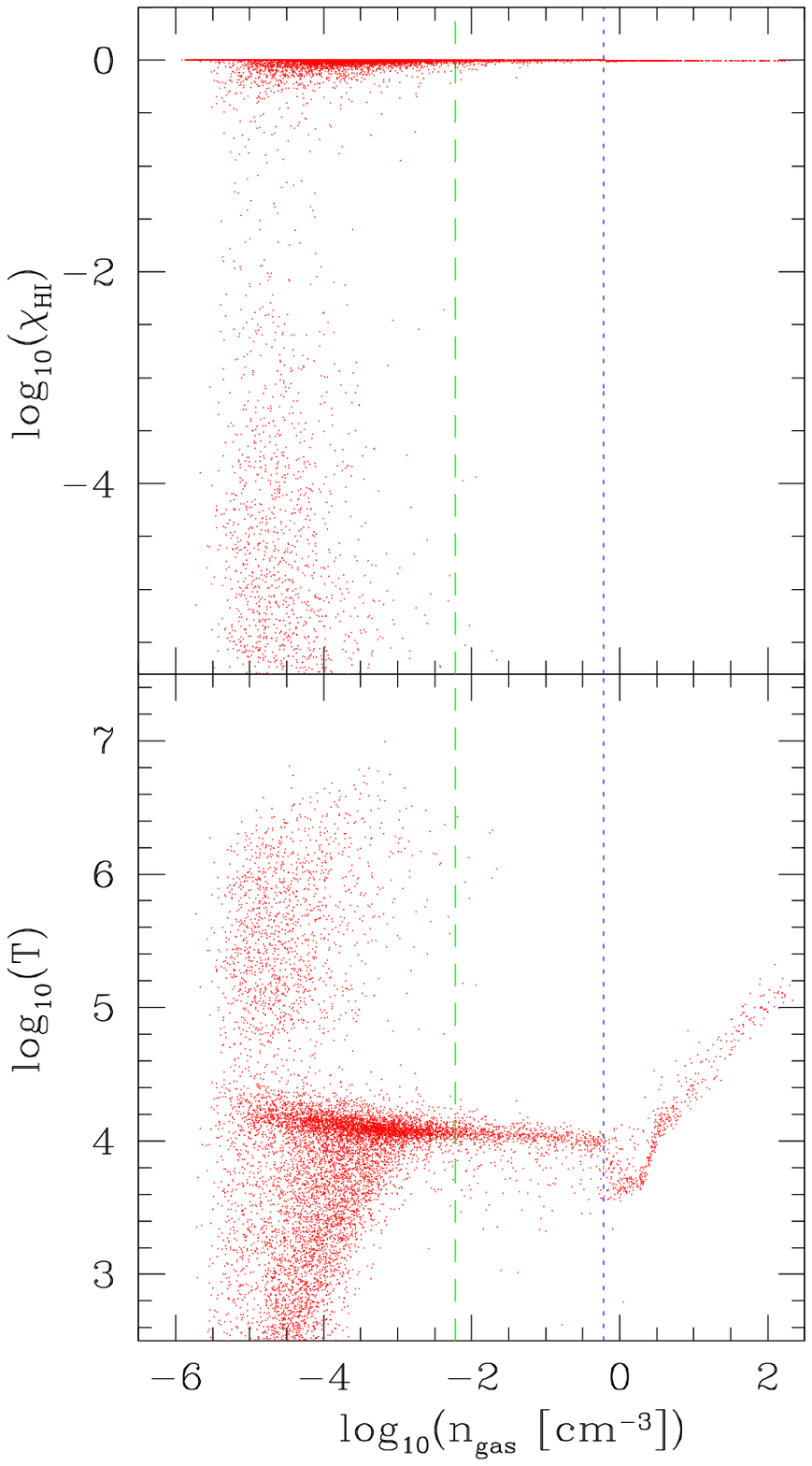}}
  \subfigure[Half-UV]{\includegraphics[width=0.225\textwidth,angle=0] {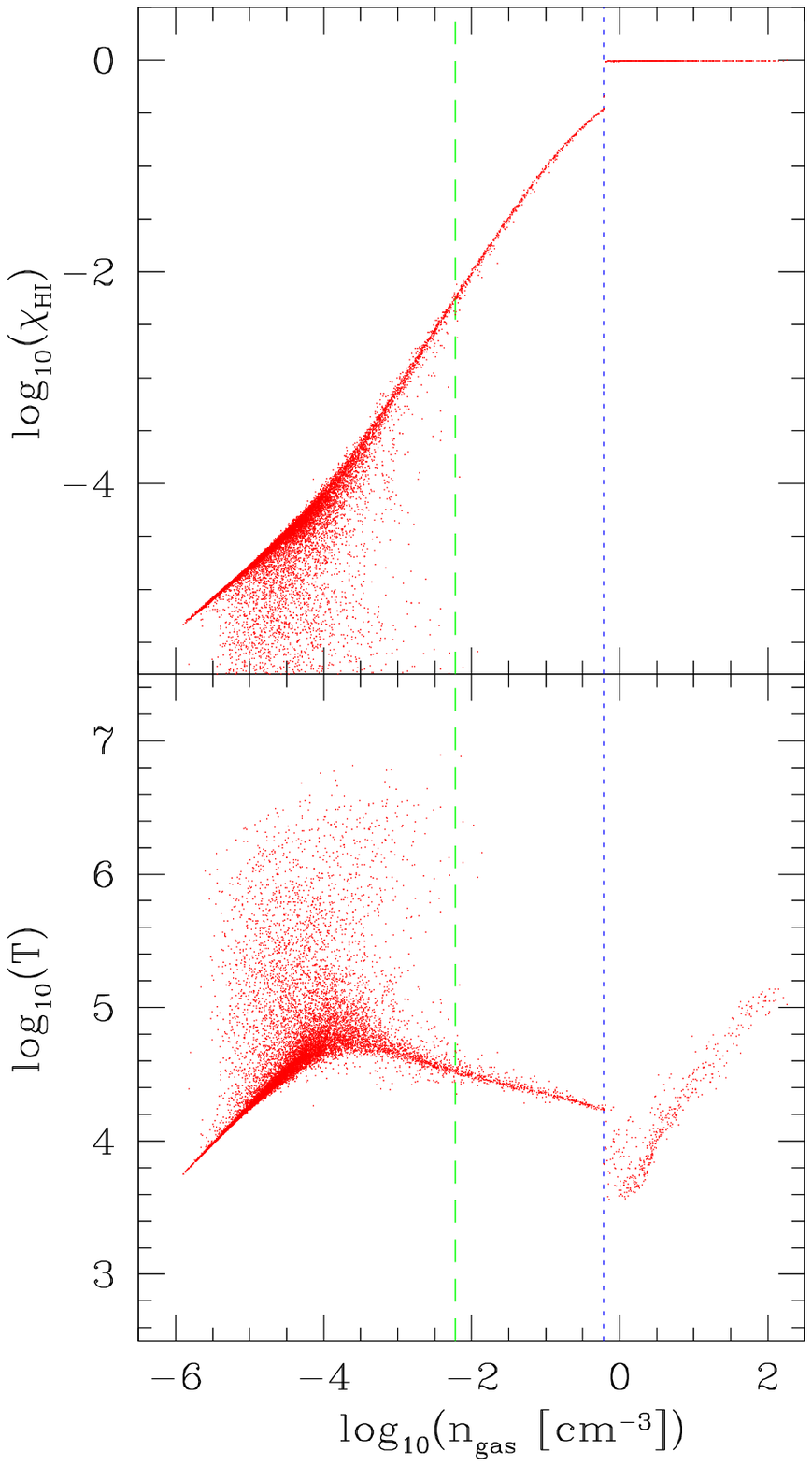}}
  \subfigure[Optically-Thick]{\includegraphics[width=0.225\textwidth,angle=0] {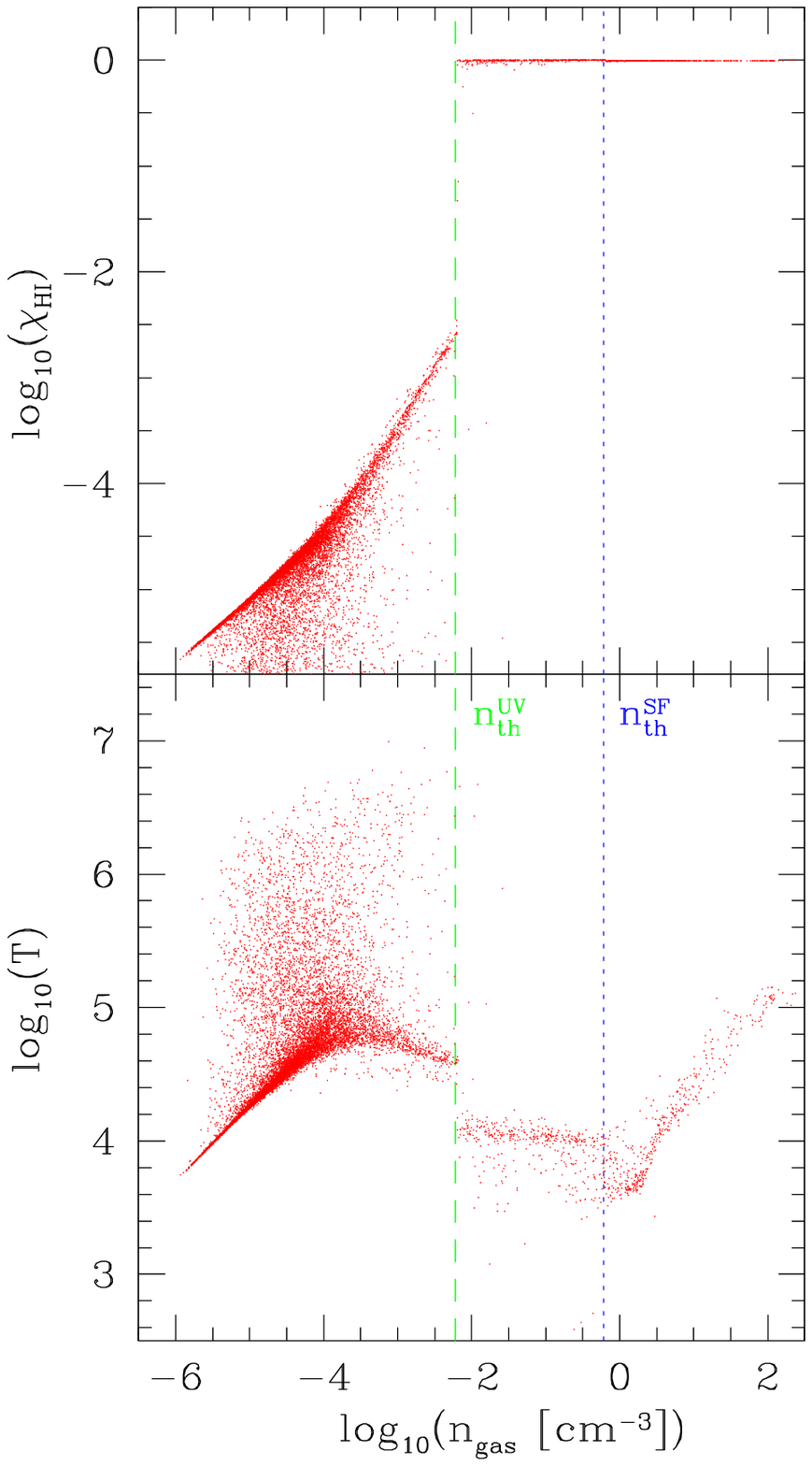}}
}
\caption{
Neutral hydrogen fraction ($\ngas$ vs. $\chi_{\rm HI}$; top panels) and phase space distribution ($\ngas$ vs. $T$; bottom panels) of the cosmic gas at $z=3$ for the four runs. The neutral fraction is defined as $\chi_{\rm HI} = n_{\rm HI} / (n_{\rm HI} + n_{\rm HII})$.  For plotting purpose, we plot only randomly selected 1\% of the total gas particles in the simulation.  
The two densities, $\nuv$ and $\nsf$, are indicated by the vertical green dashed line and blue dotted lines, respectively. 
}
\label{fig:phase}
\end{figure*}


\section{Results}
\label{sec:result}

\subsection{Effect of UVB on $\fn$}
\label{sec:uvb}

Figure\,\ref{fig:fn} shows the $\fn$ in the four runs with different 
UVB treatment, which was calculated by the same method described 
in \citet{Nag04g}. In short, we set up a uniform grid around each 
dark matter halo, and project the gas density field onto a face of 
the grid to compute $\NHI$.  
Figure\,\ref{fig:fn} clearly shows that the Fiducial run underpredicts 
the $\fn$, particularly at $\log\NHI < 21.2$.
The Half-UV run is somewhat higher than the Fiducial run, but still 
not enough to account for the observed number of columns at 
$19.5 < \log \NHI < 21$. 
On the other hand, the No-UV run completely overpredicts the observed
$\fn$ at all column densities.  
The sudden decrease in $\fn$ from the No-UV run to the Half-UV run shows 
that even a weak UVB can ionize the gas and the simulation cannot 
account for the observed number of columns. The OTUV run agrees very well 
with the observed data at $19.2<\log\NHI<21.2$.  
The comparison of these four runs suggests that the UVB affects 
the shape of $\fn$ significantly, and that applying the optically thin 
approximation at all densities is too simplistic to reproduce the 
observed $\fn$ properly.
Our simulations overpredict $\fn$ at $\log\NHI > 10^{22}$\,cm$^{-2}$. 
But this could be because our simulations do not take into account  
the conversion of \HI\ into H$_2$, and the conclusion of this paper 
does not depend on this issue. 

To obtain a better physical intuition on the effect of UVB, 
we plot the hydrogen neutral fraction ($\chiHI$) and temperature 
against gas density for the four runs in Figure~\ref{fig:phase}. 
The difference of $\fn$ between the Fiducial and the OTUV run must be 
due to the different neutral fraction of gas at $\nuv < \ngas < \nsf$
in the two runs. 
The Fiducial run predicts a high degree of ionization at 
$\ngas < \nsf$ ($\approx 0.6$\,cm$^{-3}$). 
For example, the gas with $\ngas \sim 0.1$\,cm$^{-3}$ at $z = 3$ has 
$\chiHI \sim 0.05$ for the Fiducial run.
By referencing to $\fn$, the gas with this density should correspond to 
the DLAs with $\log\NHI \sim 20$, and $\chiHI \sim 0.05$ is too low 
to match the observed data on $\fn$. 
The Half-UV run has a higher neutral fraction for the gases with 
$\ngas < \nsf$, but still with $\chiHI < 0.3$. 
In contrast, most hydrogen in the No-UV run has $\chiHI \sim 1.0$, 
except for the small fraction of hot gas which is ionized by 
collisional ionization in shocked regions. 
Eliminating the UVB completely is not a realistic assumption, 
and the No-UV run clearly overpredicts $\fn$ as expected. 
Lastly, in the OTUV run, the gas with $\ngas > \nuv$ has $\chiHI \sim 1.0$, 
which allows it to match the observed $\fn$. 
Figures\,\ref{fig:fn} and \ref{fig:phase} clearly show that the 
optically thin approximation ionizes the gas with $\nuv< \ngas < \nsf$ 
too much. 
In order to properly reproduce the observed $\fn$ with our simulations, 
the gas should be mostly neutral at $\ngas > \nuv$. 

The effects of UVB can also be seen in the $\rho-T$ phase diagrams in the bottom panels of Figure~\ref{fig:phase}. The comparison of the Fiducial and No-UV runs shows the well-known photoionization effect on the diffuse IGM at low densities of $\ngas < 10^{-2}$\,cm$^{-3}$. 
Compared to the Fiducial run, the gas with $\nuv < \ngas < \nsf$ in the OTUV run is almost fully neutral with a lower temperature of $T\sim 10^4$\,K. 
We will show in a separate paper (H. Yajima et al., 2011, in preparation)
that a full RT calculation supports the results shown in 
Figure\,\ref{fig:phase}(d). 


\subsection{Effects of Radiative Transfer on $\fn$}

The OTUV run is remarkably successful in reproducing the observed $\fn$ 
at $19<\log\NHI<21.5$, but so far we have neglected the radiative effects 
of the stellar radiation from local stellar sources, which could heat and
ionize the high-density gas in the star-forming regions. 

\begin{figure}
\begin{center}
\includegraphics[angle=0, scale=0.44]{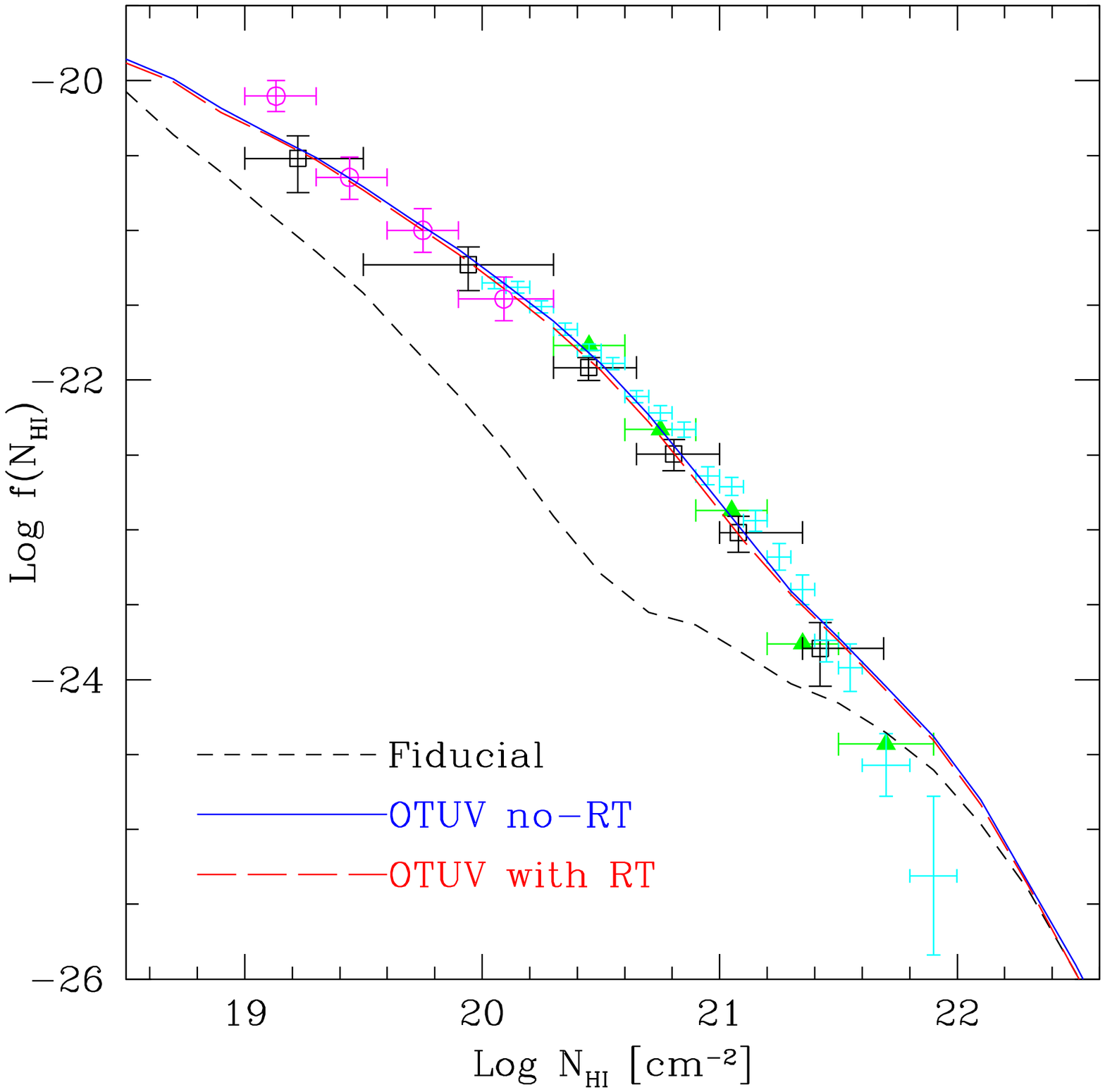}
\caption{
Effect of local stellar radiation on $\fn$ at $z=3$. 
The observational data points are the same as in Figure~\ref{fig:fn}.
The two lines for the OTUV with and without RT (red dashed and 
blue solid lines, respectively) are almost overlapping with each other. 
}
\label{fig:fn_rt}
\end{center}
\end{figure}

In order to consider the effects of local stellar radiation, 
we postprocess our simulation with the Authentic Radiation Transfer 
(ART) code \citep{Nakamoto01, Yajima09}, which uses the 
ray-tracing technique. We have performed the RT calculation in a 
similar fashion to the one we did in \citet{Yajima10}, taking 
the RT grid cell size equal to the gravitational softening length. 
This way the resolution of the RT calculation is fixed for all halos.  
Figure~\ref{fig:fn_rt} compares the results from the OTUV run 
with and without the RT calculation. 
The transferred stellar spectra are computed by using the 
{\small P{\' E}GASE} v2.0 \citep{Fioc97} based on the mass, 
formation time, and age of the star particles generated in the simulation. 
We find that the local stellar radiation does not have a 
significant impact on the shape of $\fn$, as shown by the red long-dashed 
line in Figure\,\ref{fig:fn_rt}, which is almost overlapping with 
the original OTUV run without RT. 
This is because the stellar radiation can only increase the ionization 
fraction of gas near the ionization boundary where it makes a transition
from neutral to highly ionized, and the high density neutral gas 
remains largely intact.  
We will report more details of the RT calculation results 
in a separate publication (H. Yajima et al. 2011, in preparation).


\section{Discussions and Conclusions}
\label{sec:discussion}

Using cosmological SPH simulations, we examined the effects of UVB on $\fn$, 
and clarified the reason why earlier simulations have underestimated this 
quantity compared to the observations. 
We find that the radiation sinks into the halo gas too deeply
under the optically thin approximation of UVB, and the gas 
with $19<\log\NHI<21.5$ is overly ionized. 
If we turn off the UVB above a physical density of 
$\nuv = 6\times 10^{-3}$\,cm$^{-3}$, then we reproduce the observed $\fn$ 
at $z=3$ very well. The exact value of $\nuv$ would depend on the details 
of the spectral shape and the intensity of UVB, and the size of gas cloud 
\citep{Tajiri98}.  
Based on the comparison to other works, we consider that a value of 
$\nuv \sim 10^{-2} - 10^{-3}$\,cm$^{-3}$ would be appropriate for current 
cosmological simulations. 
Our results clearly show that the optically thin approximation was 
responsible for the failure in matching the observed $\fn$ in 
earlier simulations. 

Recently, problems in the optically thin approximation have been 
pointed out in the context of the \HeII\ reionization effect on the 
thermal history of IGM \citep{McQuinn09, Faucher09} and Ly-$\alpha$ 
emission \citep{Yang06, Faucher10}. 
Our present work is another example of inadequacy of optically 
thin approximation of UVB for modeling the ionization of gas near DLAs.  
The RT effects, as well as the exact shape of UVB, 
have significant impact on the details of galaxy formation history 
including DLAs 
\citep[e.g.,][]{Tajiri98, Zheng02a, Hambrick09, Kollmeier10, Faucher10}.
                                   
Although cosmological RT simulations are beginning 
to be performed, it is still difficult to do a self-consistent 
ray-tracing RT calculation concurrently with the hydrodynamics 
due to limited computational speed. 
Under this circumstance, it would still help to have a physically 
plausible working model of self-shielding for cosmological simulations. 
Our result provides a useful proxy for the threshold density ($\nuv$) 
above which the self-shielding effect kicks in. 

We also examined the effects of local stellar radiation by postprocessing 
the simulation with a ray-tracing code. The effect is not as strong
as the UVB, and it only slightly decreases $\fn$. 
We will report the further details of this ray-tracing work and its 
effect on DLA cross section in a separate publication 
(H. Yajima et al. 2011, in preparation). 

To put it in another way, our results suggest that the shape of $\fn$ 
and the ionization structure of sub-DLAs and DLAs would be great probes 
of UVB at high redshift, and that the DLA gas is closely related to 
the transition region from optically-thin ionized gas to 
optically-thick neutral gas within dark matter halos. 
Further comparisons on the physical properties of DLAs, sub-DLAs, 
and Lyman limit systems between simulations and observations would 
give us useful insight on the nature of UVB. 

In Section~\ref{sec:sim} we argued that our results are not strongly 
affected by the numerical resolution of our current simulations, 
based on the results of \citet{Nag04g}. However the simulations of 
\citet{Nag04g} did not include the self-shielding treatment of the 
OTUV run, so to really test the resolution effect, we would have to 
run a new OTUV simulation with $2\times 324^3$ particles, which we plan to 
perform in the near future and address the resolution effects 
more directly. 


\section*{Acknowledgments}

This work is supported in part by the NSF Grant AST-0807491, 
National Aeronautics and Space Administration under Grant/Cooperative 
Agreement No. NNX08AE57A issued by the Nevada NASA EPSCoR program, and 
the President's Infrastructure Award from UNLV. 
KN is grateful for the hospitality of the IPMU at University of Tokyo, 
where part of this work was performed.



\begin{thebibliography}{34}
\expandafter\ifx\csname natexlab\endcsname\relax\def\natexlab#1{#1}\fi

\bibitem[{{Choi} \& {Nagamine}(2009{\natexlab{a}})}]{Choi09b}
{Choi}, J. \& {Nagamine}, K. 2009{\natexlab{a}}, \mnras, 393, 1595

\bibitem[{{Choi} \& {Nagamine}(2009{\natexlab{b}})}]{Choi09a}
---. 2009{\natexlab{b}}, \mnras, 395, 1776

\bibitem[{{Choi} \& {Nagamine}(2010)}]{Choi10a}
---. 2010, \mnras, 407, 1464

\bibitem[{{Dav{\'e}} {et~al.}(1999){Dav{\'e}}, {Hernquist}, {Katz}, \&
  {Weinberg}}]{Dave99}
{Dav{\'e}}, R., {Hernquist}, L., {Katz}, N., \& {Weinberg}, D.~H. 1999, \apj,
  511, 521

\bibitem[{{Faucher-Giguere} {et~al.}(2010){Faucher-Giguere}, {Keres},
  {Dijkstra}, {Hernquist}, \& {Zaldarriaga}}]{Faucher10}
{Faucher-Giguere}, C., {Keres}, D., {Dijkstra}, M., {Hernquist}, L., \&
  {Zaldarriaga}, M. 2010, arXiv e-prints (arXiv:1005.3041)

\bibitem[{{Faucher-Gigu{\`e}re} {et~al.}(2009){Faucher-Gigu{\`e}re}, {Lidz},
  {Zaldarriaga}, \& {Hernquist}}]{Faucher09}
{Faucher-Gigu{\`e}re}, C., {Lidz}, A., {Zaldarriaga}, M., \& {Hernquist}, L.
  2009, \apj, 703, 1416

\bibitem[{{Fioc} \& {Rocca-Volmerange}(1997)}]{Fioc97}
{Fioc}, M. \& {Rocca-Volmerange}, B. 1997, \aap, 326, 950

\bibitem[{Haardt \& Madau(1996)}]{Haardt96}
Haardt, F. \& Madau, P. 1996, ApJ, 461, 20

\bibitem[{{Hambrick} {et~al.}(2009){Hambrick}, {Ostriker}, {Naab}, \&
  {Johansson}}]{Hambrick09}
{Hambrick}, D.~C., {Ostriker}, J.~P., {Naab}, T., \& {Johansson}, P.~H. 2009,
  \apj, 705, 1566

\bibitem[{Katz {et~al.}(1996{\natexlab{a}})Katz, Weinberg, \&
  Hernquist}]{Katz96a}
Katz, N., Weinberg, D.~H., \& Hernquist, L. 1996{\natexlab{a}}, ApJS, 105, 19

\bibitem[{Katz {et~al.}(1996{\natexlab{b}})Katz, Weinberg, Hernquist, \&
  Miralda-Escud\'e}]{Katz96b}
Katz, N., Weinberg, D.~H., Hernquist, L., \& Miralda-Escud\'e, J.
  1996{\natexlab{b}}, ApJ, 457, L57

\bibitem[{{Kollmeier} {et~al.}(2010){Kollmeier}, {Zheng}, {Dav{\'e}}, {Gould},
  {Katz}, {Miralda-Escud{\'e}}, \& {Weinberg}}]{Kollmeier10}
{Kollmeier}, J.~A., {Zheng}, Z., {Dav{\'e}}, R., {Gould}, A., {Katz}, N.,
  {Miralda-Escud{\'e}}, J., \& {Weinberg}, D.~H. 2010, \apj, 708, 1048

\bibitem[{{Komatsu} {et~al.}(2009){Komatsu}, {Dunkley}, {Nolta}, {Bennett},
  {et~al.}}]{Komatsu09}
{Komatsu}, E., {et~al.} 2009, \apjs, 180, 330

\bibitem[{{Komatsu} {et~al.}(2010){Komatsu}, {Smith}, {Dunkley}, {Bennett},
  {et~al.}}]{Komatsu10}
{Komatsu}, E., {et~al.} 2010, arXiv e-prints (arXiv:1001.4538)

\bibitem[{{McQuinn} {et~al.}(2009){McQuinn}, {Lidz}, {Zaldarriaga},
  {Hernquist}, {Hopkins}, {Dutta}, \& {Faucher-Gigu{\`e}re}}]{McQuinn09}
{McQuinn}, M., {Lidz}, A., {Zaldarriaga}, M., {Hernquist}, L., {Hopkins},
  P.~F., {Dutta}, S., \& {Faucher-Gigu{\`e}re}, C. 2009, \apj, 694, 842

\bibitem[{{Nagamine} {et~al.}(2004){Nagamine}, {Springel}, \&
  {Hernquist}}]{Nag04g}
{Nagamine}, K., {Springel}, V., \& {Hernquist}, L. 2004, \mnras, 348, 421

\bibitem[{{Nagamine} {et~al.}(2007){Nagamine}, {Wolfe}, {Hernquist}, \&
  {Springel}}]{Nag07a}
{Nagamine}, K., {Wolfe}, A.~M., {Hernquist}, L., \& {Springel}, V. 2007, \apj,
  660, 945

\bibitem[{{Nakamoto} {et~al.}(2001){Nakamoto}, {Umemura}, \&
  {Susa}}]{Nakamoto01}
{Nakamoto}, T., {Umemura}, M., \& {Susa}, H. 2001, \mnras, 321, 593

\bibitem[{{Noterdaeme} {et~al.}(2009){Noterdaeme}, {Petitjean}, {Ledoux}, \&
  {Srianand}}]{Noterdaeme09}
{Noterdaeme}, P., {Petitjean}, P., {Ledoux}, C., \& {Srianand}, R. 2009, \aap,
  505, 1087

\bibitem[{{O'Meara} {et~al.}(2007){O'Meara}, {Prochaska}, {Burles}, {Prochter},
  {Bernstein}, \& {Burgess}}]{Omeara07}
{O'Meara}, J.~M., {Prochaska}, J.~X., {Burles}, S., {Prochter}, G.,
  {Bernstein}, R.~A., \& {Burgess}, K.~M. 2007, \apj, 656, 666

\bibitem[{{P{\'e}roux} {et~al.}(2005){P{\'e}roux}, {Dessauges-Zavadsky},
  {D'Odorico}, {Sun Kim}, \& {McMahon}}]{Peroux05}
{P{\'e}roux}, C., {Dessauges-Zavadsky}, M., {D'Odorico}, S., {Sun Kim}, T., \&
  {McMahon}, R.~G. 2005, \mnras, 363, 479

\bibitem[{{P{\'e}roux} {et~al.}(2003){P{\'e}roux}, {McMahon},
  {Storrie-Lombardi}, \& {Irwin}}]{Peroux03}
{P{\'e}roux}, C., {McMahon}, R.~G., {Storrie-Lombardi}, L.~J., \& {Irwin},
  M.~J. 2003, \mnras, 346, 1103

\bibitem[{{Prochaska} {et~al.}(2008){Prochaska}, {Hennawi}, \&
  {Herbert-Fort}}]{Pro08c}
{Prochaska}, J.~X., {Hennawi}, J.~F., \& {Herbert-Fort}, S. 2008, \apj, 675,
  1002

\bibitem[{{Prochaska} \& {Herbert-Fort}(2004)}]{Pro04c}
{Prochaska}, J.~X. \& {Herbert-Fort}, S. 2004, \pasp, 116, 622

\bibitem[{{Prochaska} {et~al.}(2005){Prochaska}, {Herbert-Fort}, \&
  {Wolfe}}]{Pro05}
{Prochaska}, J.~X., {Herbert-Fort}, S., \& {Wolfe}, A.~M. 2005, ApJ, 635, 123

\bibitem[{{Prochaska} \& {Wolfe}(2009)}]{Pro09}
{Prochaska}, J.~X. \& {Wolfe}, A.~M. 2009, \apj, 696, 1543

\bibitem[{{Schaye} \& {Dalla Vecchia}(2008)}]{Schaye08}
{Schaye}, J. \& {Dalla Vecchia}, C. 2008, \mnras, 383, 1210

\bibitem[{{Springel}(2005)}]{Springel05e}
{Springel}, V. 2005, \mnras, 364, 1105

\bibitem[{{Springel} \& {Hernquist}(2003)}]{Springel03b}
{Springel}, V. \& {Hernquist}, L. 2003, \mnras, 339, 289

\bibitem[{{Tajiri} \& {Umemura}(1998)}]{Tajiri98}
{Tajiri}, Y. \& {Umemura}, M. 1998, \apj, 502, 59

\bibitem[{{Yajima} {et~al.}(2010){Yajima}, {Choi}, \& {Nagamine}}]{Yajima10}
{Yajima}, H., {Choi}, J., \& {Nagamine}, K. 2010, MNRAS, in press (arXiv:1002.3346)

\bibitem[{{Yajima} {et~al.}(2009){Yajima}, {Umemura}, {Mori}, \&
  {Nakamoto}}]{Yajima09}
{Yajima}, H., {Umemura}, M., {Mori}, M., \& {Nakamoto}, T. 2009, \mnras, 398,
  715

\bibitem[{{Yang} {et~al.}(2006){Yang}, {Zabludoff}, {Dav{\'e}}, {Eisenstein},
  {Pinto}, {Katz}, {Weinberg}, \& {Barton}}]{Yang06}
{Yang}, Y., {Zabludoff}, A.~I., {Dav{\'e}}, R., {Eisenstein}, D.~J., {Pinto},
  P.~A., {Katz}, N., {Weinberg}, D.~H., \& {Barton}, E.~J. 2006, \apj, 640, 539

\bibitem[{{Zheng} \& {Miralda-Escud{\'e}}(2002)}]{Zheng02a}
{Zheng}, Z. \& {Miralda-Escud{\'e}}, J. 2002, ApJ, 568, L71

\end{thebibliography}

\end{document}